\documentclass[11pt]{article}
\usepackage{moriond}
\usepackage{ifthen}
\usepackage{graphicx}

\def\ra{\rightarrow}

\def\be{\begin{equation}}
\def\ee{\end{equation}}
\def\bea{\begin{eqnarray}}
\def\eea{\end{eqnarray}}

\newcommand{\pt}{\ensuremath{{p_T}}}
\newcommand{\ptmu}{\ensuremath{{p_{T,\mu}}}}
\newcommand{\ptrel}{\ensuremath{{p_T^{rel}}}}
\newcommand{\call}{{\cal L}}
\newcommand{\ilumi}{\ensuremath{\mathrm{cm^{-2}s^{-1}}}}
\newcommand{\ipb}{\ensuremath{\mathrm{pb^{-1}}}}
\newcommand{\ifb}{\ensuremath{\mathrm{fb^{-1}}}}
\newcommand{\ccbar}{\ensuremath{{c\bar{c}}}}
\newcommand{\bbbar}{\ensuremath{{b\bar{b}}}}
\newcommand{\sbbbar}{\ensuremath{{\sigma_\bbbar}}}

\newcommand{\jpsi}{\ensuremath{\mathrm{J/\psi}}}
\newcommand{\sjpsi}{\ensuremath{{\sigma_{\jpsi}}}}
\newcommand{\supsilon}{\ensuremath{{\sigma_{\Upsilon}}}}
\newcommand{\dsd}[2]{\ensuremath{\frac{d\sigma_{#1}}{d #2}}}
\newcommand{\bplusjpsik}{{\ensuremath{B^+\ra\jpsi\,K^+}}}

\newcommand{\Upsilontomumu}{\ensuremath{\Upsilon\ra\mu^+\mu^-}}

\renewcommand{\authorphoto}{%
  \includegraphics[height=35mm]{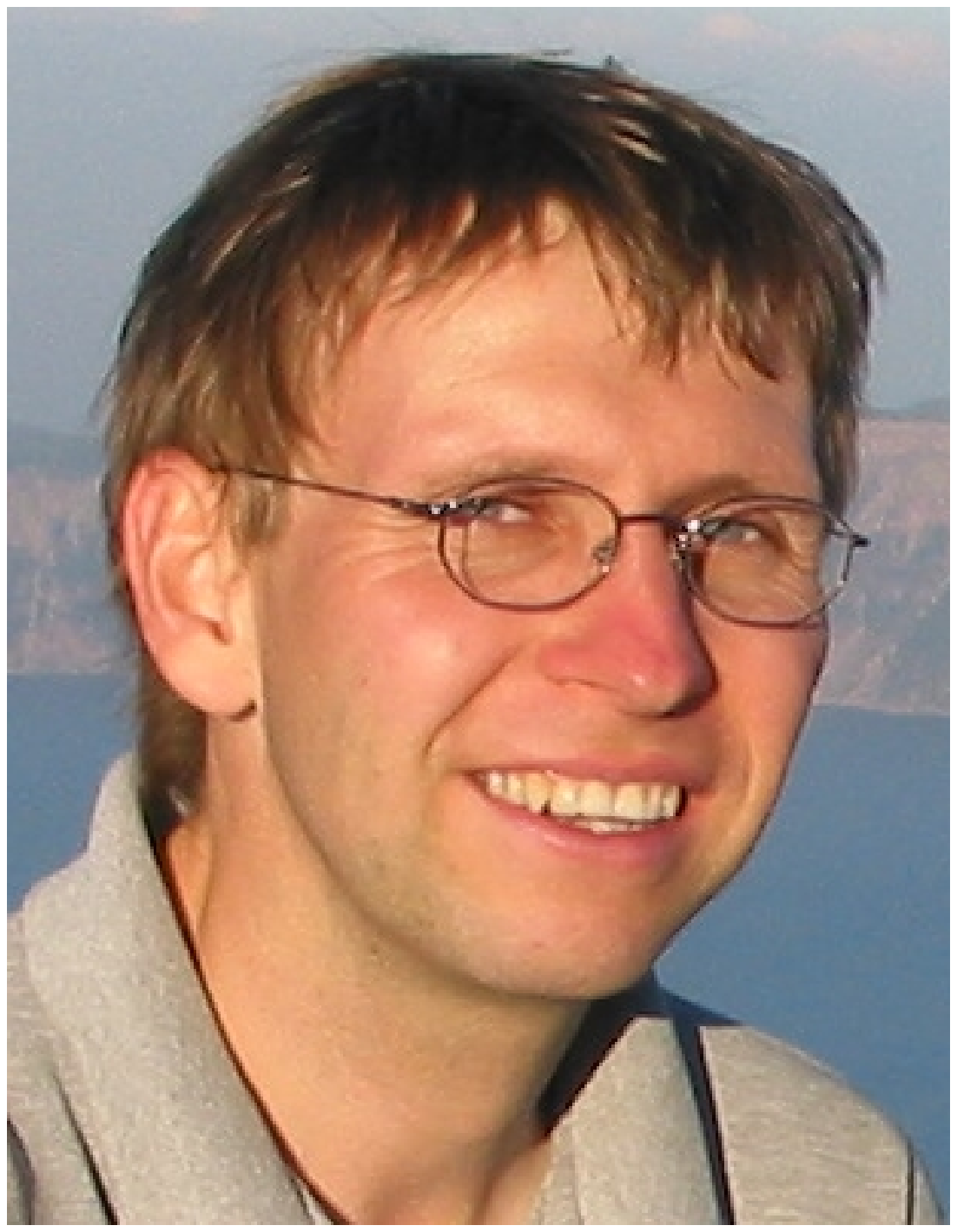}
  }

\newboolean{draft}
\setboolean{draft}{false}

\newboolean{docomments}
\setboolean{docomments}{true}
\newcommand{\comment}[1]{\protect 
  \ifthenelse{\boolean{docomments}}{{\em #1}}{}}

\ifthenelse{\boolean{draft}}{}{{\setboolean{docomments}{false}}}

\newcommand{\svnrev}{32}

\begin{document}
\ifthenelse{\boolean{draft}}{
  \begin{flushright}
    Draft SVN rev. \svnrev\ -- \today
  \end{flushright}
}{
}
\begin{flushright}
  SI-HEP-2008-09\\
  ATL-PHYS-PROC-2008-033
\end{flushright}
\vspace*{2.5cm}

\title{Charm and Bottom Production Measurements at the LHC}

\author{Wolfgang Walkowiak\\
        on behalf of the ATLAS and CMS collaborations}

\address{University of Siegen, 
         Department of Physics, Experimental Particle Physics, \\
         57068 Siegen, Germany\ \footnote{%
          Email: wolfgang.walkowiak@cern.ch}}

\maketitle

\abstracts{Early data of the ATLAS and CMS experiments at the
  LHC will allow us to measure the cross sections for beauty and
  heavy quarkonia production in $pp$ collisions
  at a center-of-mass energy of 14~TeV to a reasonable
  precision.  Different experimental approaches employing single or 
  di-muon triggered events and $b$-tagging methods are 
  discussed.  The potential for extracting the polarization of 
  vector states from the decays $\jpsi\ra \mu^+\mu^-$ and 
  $\Upsilon\ra \mu^+\mu^-$ is presented. 
}

\section{Introduction}

The Large Hadron Collider (LHC) at CERN is expected to start operating
in 2008. 
With an expected \bbbar\ cross section \sbbbar\ 
of approximately 500~$\mu$b,
corresponding to one \bbbar\ event in 100~proton-proton collisions, 
the LHC provides about $2\times10^{12}$ $b\bar{b}$ per year (at 
$\call = 10^{33}$~\ilumi).  However, due to uncertainties in the
extrapolation of \sbbbar\ from measurements at
lower energies being as large as a factor two, it is necessary to
measure \sbbbar\ 
as well as differential cross sections as functions of \pt\ and $\eta$
with early LHC data.

The measurement of the production cross sections \sjpsi\ and
\supsilon\ for prompt heavy quarkonia\,\cite{kartvelishvili}, 
i.e. bound states of \ccbar\ and \bbbar\ quarks, 
provide a good testbed for various QCD models.  
The theoretical description\,\cite{kramer-cdf} 
of the measured excess, e.g. in the 
direct $\jpsi\ra\mu^+\mu^-$ cross section \dsd{\jpsi}{\pt}\ at 
CDF\,\cite{Abe:CDF}, 
required a contribution in addition to the 
color singlet model
(CSM).  Although the color octet 
model
(COM),
which includes an evolution of the quark-antiquark quantum
mechanical state produced to the bound quarkonium state, describes
\dsd{\jpsi,\Upsilon}{\pt}\ well,
some predictions of the COM for the polarization 
of quarkonia decays such as \Upsilontomumu\ are very 
different from Tevatron data\,\cite{D0polar}.

The trigger strategies for $B$ physics and quarkonia at ATLAS and CMS
rely on single and di-muon trigger elements with low \pt\ thresholds.  
At Level~1, implemented in hardware, single muon trigger thresholds
from $\ptmu > 4$~GeV upwards and di-muon \pt\ thresholds as low as 
$\ptmu > 3$~GeV will be employed, depending on the instantaneous
luminosity.  The software-implemented higher level trigger algorithms 
(HLT) will partially reconstruct exclusive $B$ decays 
involving e.g. $D_s\ra \phi \pi$
and \jpsi\ to $e/\gamma$ or $\mu^+\mu^-$ final states.  CMS plans to
implement an inclusive trigger for charm and bottom decays using
b-tagging algorithms.  At ATLAS the single muon trigger rate is
expected to reach approximately 10~kHz (with $\ptmu > 6$~GeV at 
$\call = 10^{33}$~\ilumi) with the di-muon trigger rate being about two
orders of magnitude lower.  Prompt $\jpsi\ra\mu^+\mu^-$ production
contributes at the level of a few Hertz only. 

\begin{figure}[tb]
  \begin{center}
    \parbox{0.48\textwidth}{
      \includegraphics[width=0.5\textwidth]{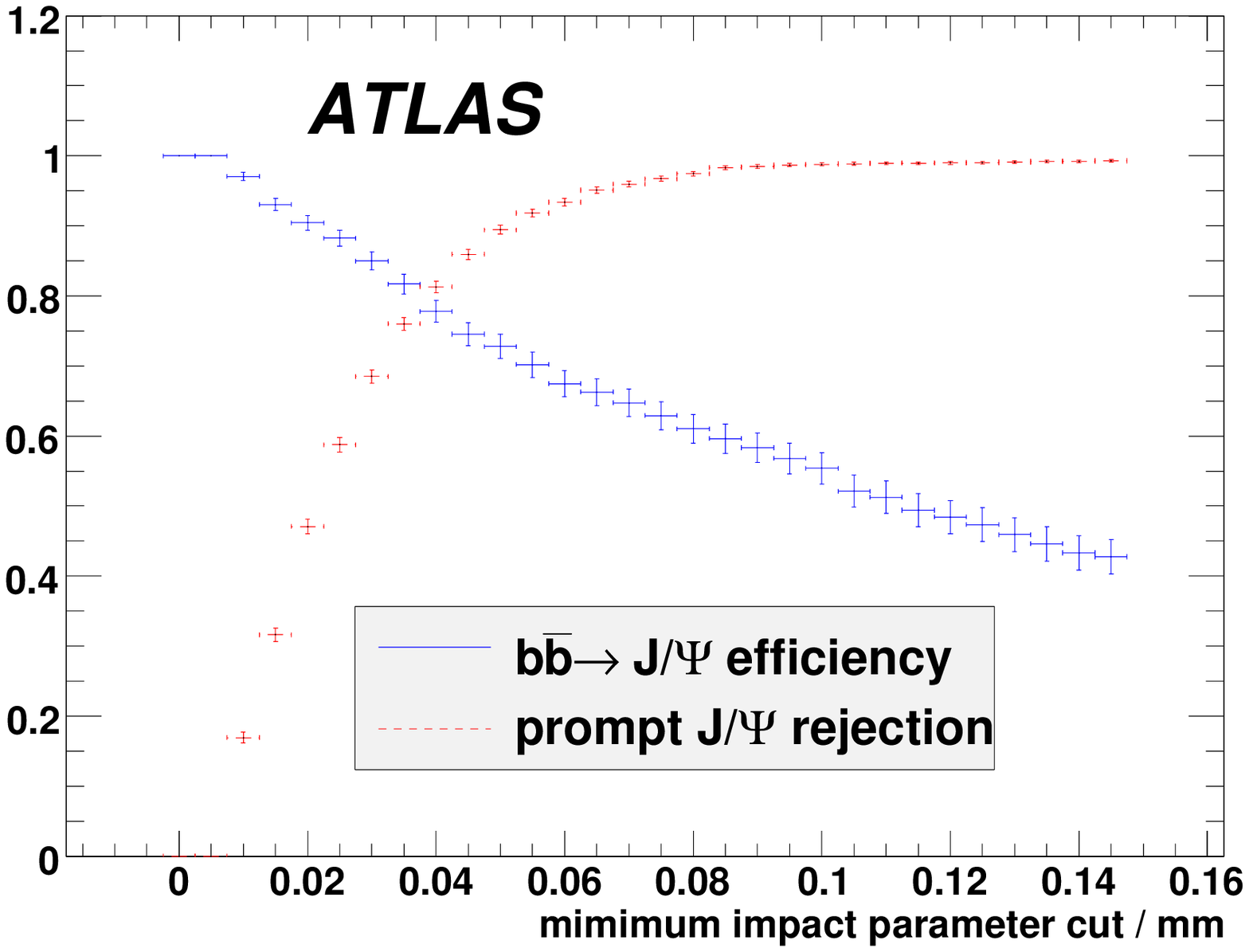}
      \vspace*{-0.8cm}
       \caption{\label{fig:bcrossipcut}%
	 The Monte Carlo $\bbbar\ra\jpsi(\mu^+\mu^-)$ signal
	 efficiency (blue falling data points) and the rejection 
         of prompt $\jpsi(\mu^+\mu^-)$ events (red rising data points)
         as a function of the minimum impact parameter cut for the
	 ATLAS study.
       }
    } 
    \hfill
    \parbox{0.48\textwidth}{
      \includegraphics[width=0.5\textwidth]{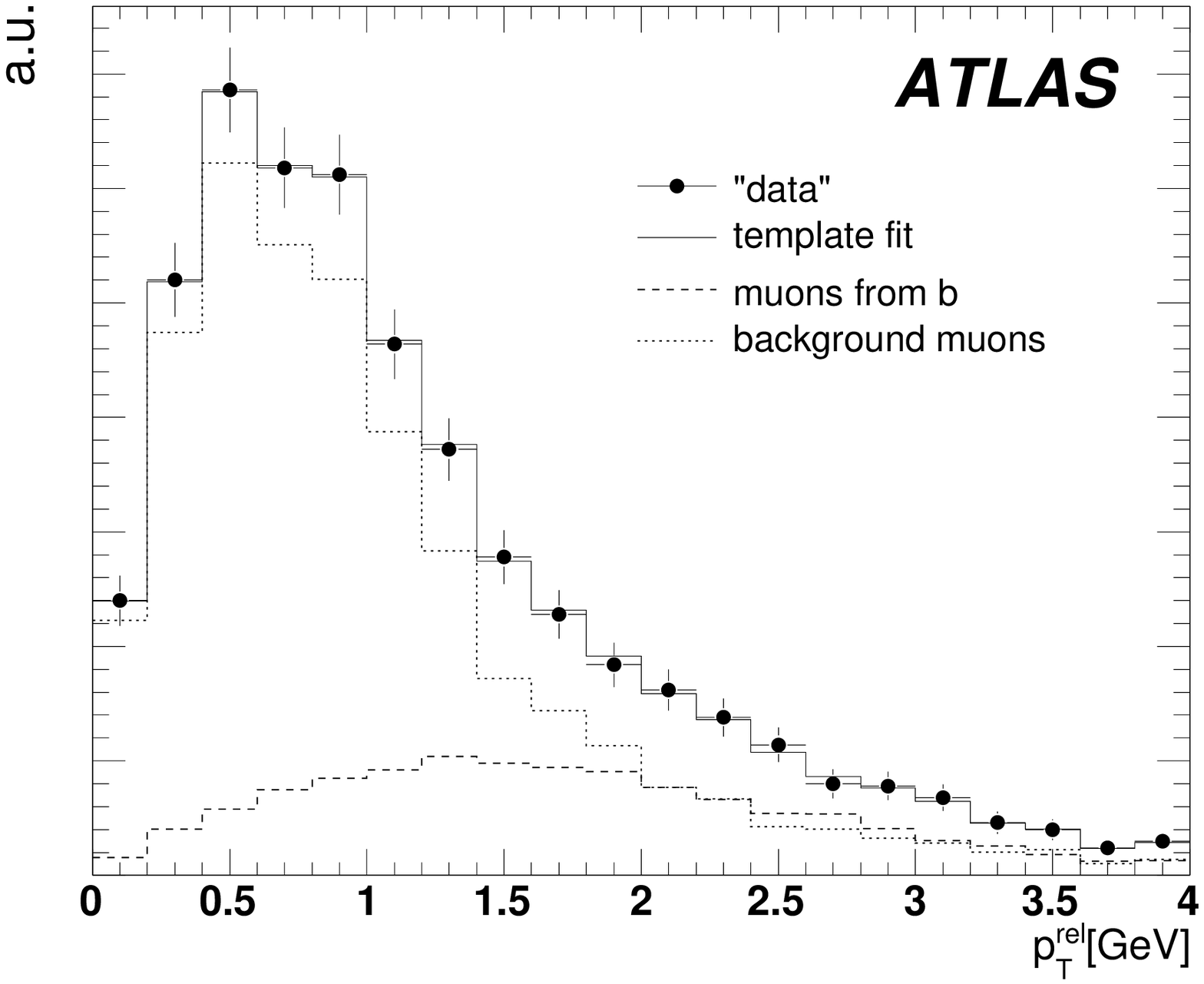}
      \vspace*{-0.8cm}
      \caption{\label{fig:bcrosstempfit}%
	The test fit of shape templates to the Monte Carlo data
	sample results in a $b$-fraction of $(23\pm2)$\% and a 
	corresponding background fraction of $(77\pm4)$\% for 
        the ATLAS study.
      }
    }
  \end{center}
\end{figure}

\section{b Cross Section Measurements}


Two methods for the measurement of the inclusive $b$ cross section
at low \pt\ 
have been studied by ATLAS
for the early data taking period.  
The first method considered employs a di-muon $\jpsi$ trigger 
($p_{T,\mu} > 4$~GeV), applies offline cuts of 6 and 4~GeV $p_T$
to the two identified muons and 
reconstructs the $\jpsi\ra\mu^+\mu^-$ signature requiring a detached 
\jpsi\ vertex.  
Requesting a minimum
impact parameter of the leading muon of 0.08~mm results in an
efficiency of about 60\% for the $\bbbar\ra \jpsi\,X$ signal decays
while almost all prompt \jpsi\ events are rejected 
(Fig.~\ref{fig:bcrossipcut}). 
The second method is based on semileptonic $b\ra \mu$ decays.
A single muon ($p_{T,\mu} > 6$~GeV) plus a jet
region-of-interest is required (13.5\% trigger efficiency).
Events containing $b$-induced jets  are enriched 
using $b$-jet weight tagging and
exploiting the fact that the relative momentum of an associated muon 
w.r.t. the jet axis \ptrel\ 
is on average larger for jets containing a $b$ particle than for other jets.
The signal reconstruction efficiency amounts to 85\%.
A fit of the simulated data to shape templates in \ptrel,
which are determined from Monte Carlo,
is used to extract the relative fractions of signal
($b\ra\mu$ and $b\ra c\ra\mu$) and background ($c\ra\mu$ and $\pi/K\ra
\mu$) events (Fig.~\ref{fig:bcrosstempfit}).

A statistical
precision of the \bbbar\ cross section measurement of ${\cal O}(1\%)$
after typically one month of data taking is expected (at  
$\call = 10^{31}$~\ilumi\ or higher), 
with estimated systematic uncertainties of
approximately 9\% for 300~\ipb.  


In an ATLAS study for the 
channel \bplusjpsik,
a di-muon \jpsi\ trigger is used (82\% efficiency).
The offline algorithms reconstruct \jpsi\ candidates ($p_{T,\mu_1} > 6$~GeV, 
$p_{T,\mu_2} > 3$~GeV) requiring a displaced vertex ($\lambda >
100~\mu$m) with an efficiency of 55.8\% to reduce combinatorial background
from prompt \jpsi.
Adding an additional track
with $\pt > 1.5$~GeV and a large impact parameter, the $B^+$
candidates are selected by cuts on the vertex displacement ($\lambda >
100~\mu$m) and on a mass window of $\pm120$~GeV around the nominal
$B^+$ mass.  An overal signal
reconstruction efficiency of $(29.8\pm0.84)$\% and a $B^+$ mass
resolution of $(42.2\pm1.3)$~MeV are obtained.
An early data sample of 
e.g.\ 13.2~\ipb, corresponding to about 2\,100 signal events, 
will allow us to measure 
the cross section and mass resolution to about 3\%
and the $B^+$ lifetime to about 2\% statistical precision.

\begin{figure}[tb]
  \begin{center}
    \parbox{0.48\textwidth}{
      \includegraphics[width=0.45\textwidth]{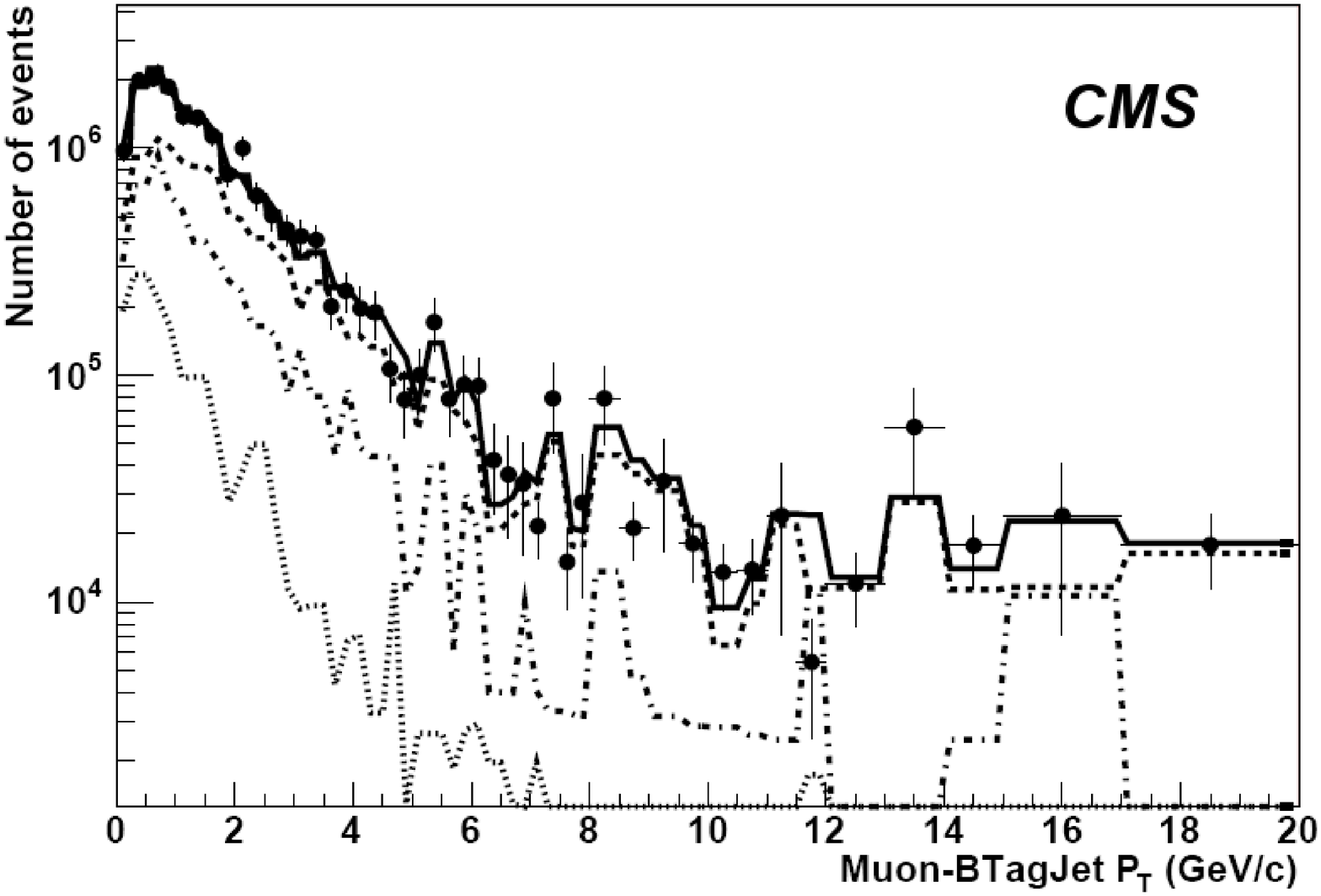}
      \vspace*{-0.3cm}%
      \caption{\label{fig:bjetcrosstempfit}%
	Fit of the muon \ptrel\ spectrum 
        for the CMS Monte Carlo
	study\,\protect\cite{CMSbjet} for 10~\ifb.  
        The contributions, as a result of the shape template fit,
	from $b$ (dashed curve), $c$ (dot-dashed curve) 
	and light quark events (dotted curve) are shown, 
        as well as the total sum.
      }
    }
    \hfill
    \parbox{0.48\textwidth}{
      \includegraphics[width=0.45\textwidth]{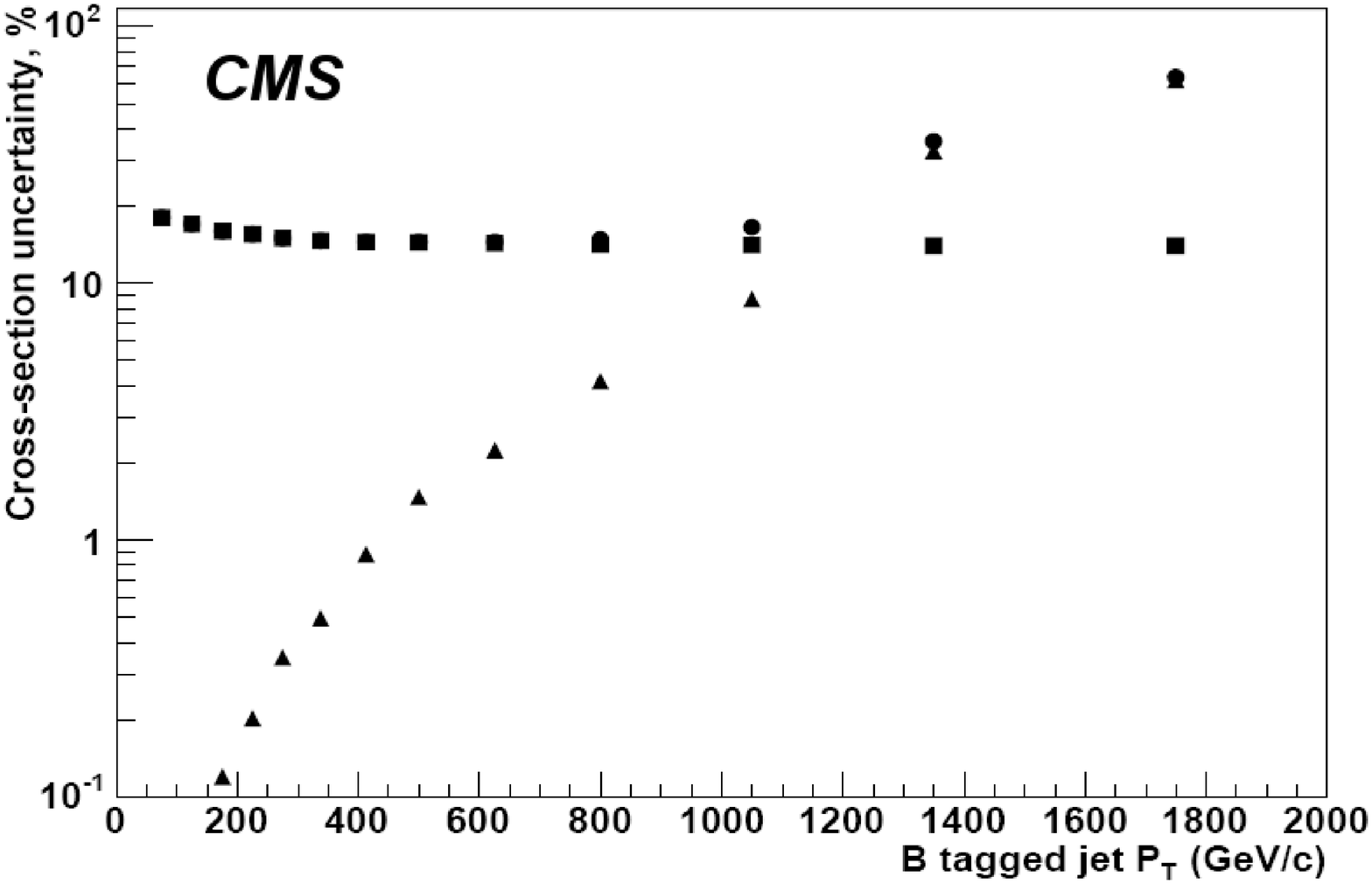}
      \caption{\label{fig:bjetcrosserrors}%
        The statistical (triangles), systematic (squares) and total
        (circles) uncertainty for the $b$ jet cross section
        measurement as a function of the $b$ tagged jet \pt\ for the
        CMS study\,\protect\cite{CMSbjet}.
      }
    }     
  \end{center}
\end{figure}


The CMS study for an inclusive b-jet cross section
measurement\,\cite{CMSbjet} requires a trigger 
on a single muon of $p_{T,\mu}>14$~GeV
within $|\eta| < 2.1$ at Level 1 (18\% efficiency) and a muon
of $p_{T,\mu}>19$~GeV and a jet with $E_{T,jet} > 50$~GeV within
$|\eta| < 2.4$ at the HLT (60\% efficiency).  
The offline selection combines a b-tagged jet with similar properties, 
reconstructed with 65\% (55\%) efficiency in the barrel (endcap),
with a muon associated with this jet.  
The total efficiency of about 5\% results in
16~million \bbbar\ events per 10~\ifb.
Again, the relative momentum of the muon w.r.t. the $b$-jet axis
\ptrel\ is used in order to extract the $b$-jet fraction from a shape
template fit, see Fig.~\ref{fig:bjetcrosstempfit}.  The light-quark
background shape will be derived directly from data by applying an
anti-b-tag selection, while the shapes of the $b$ and $c$
contributions will be taken from Monte Carlo.  
The $b$-jet purity varies from 70\% to 55\% as the jet \pt\
increases.
We expect to observe $B$ hadrons with a \pt\ up to 1.5~TeV.
Fig.~\ref{fig:bjetcrosserrors} displays the estimated statistical,
systematic and total uncertainty as a function of the $b$-tagged jet
\pt.  The systematic uncertainties are dominated by uncertainties
in the jet energy scale (12\%) and in the fragmentation modelling
(9\%).

\section{Heavy Quarkonia}

\begin{figure}[tb]
  \begin{center}
    \parbox{0.48\textwidth}{
      \includegraphics[width=0.5\textwidth]{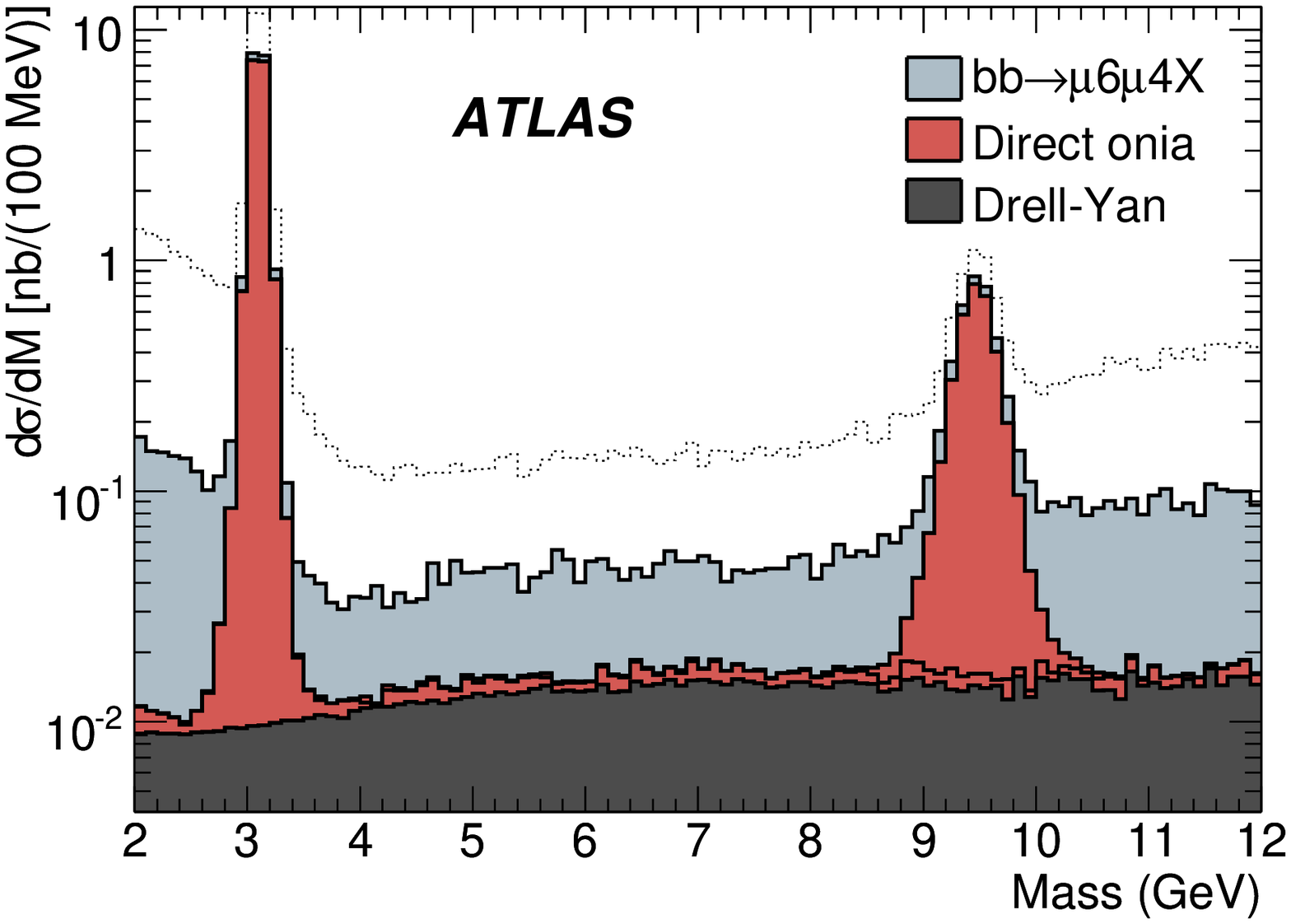}
      \vspace*{-1.0cm}%
      \caption{\label{fig:quarkoniamasses}%
	Sources of low invariant mass di-muon events, reconstructed
	with a di-muon trigger for the ATLAS study.
        Both muons are required to originate from the primary vertex.
	The light dotted line highlights the background level
	   before vertexing cuts.
      }
    }
    \hfill
    \parbox{0.48\textwidth}{
      \centering
      \includegraphics[width=0.37\textwidth]{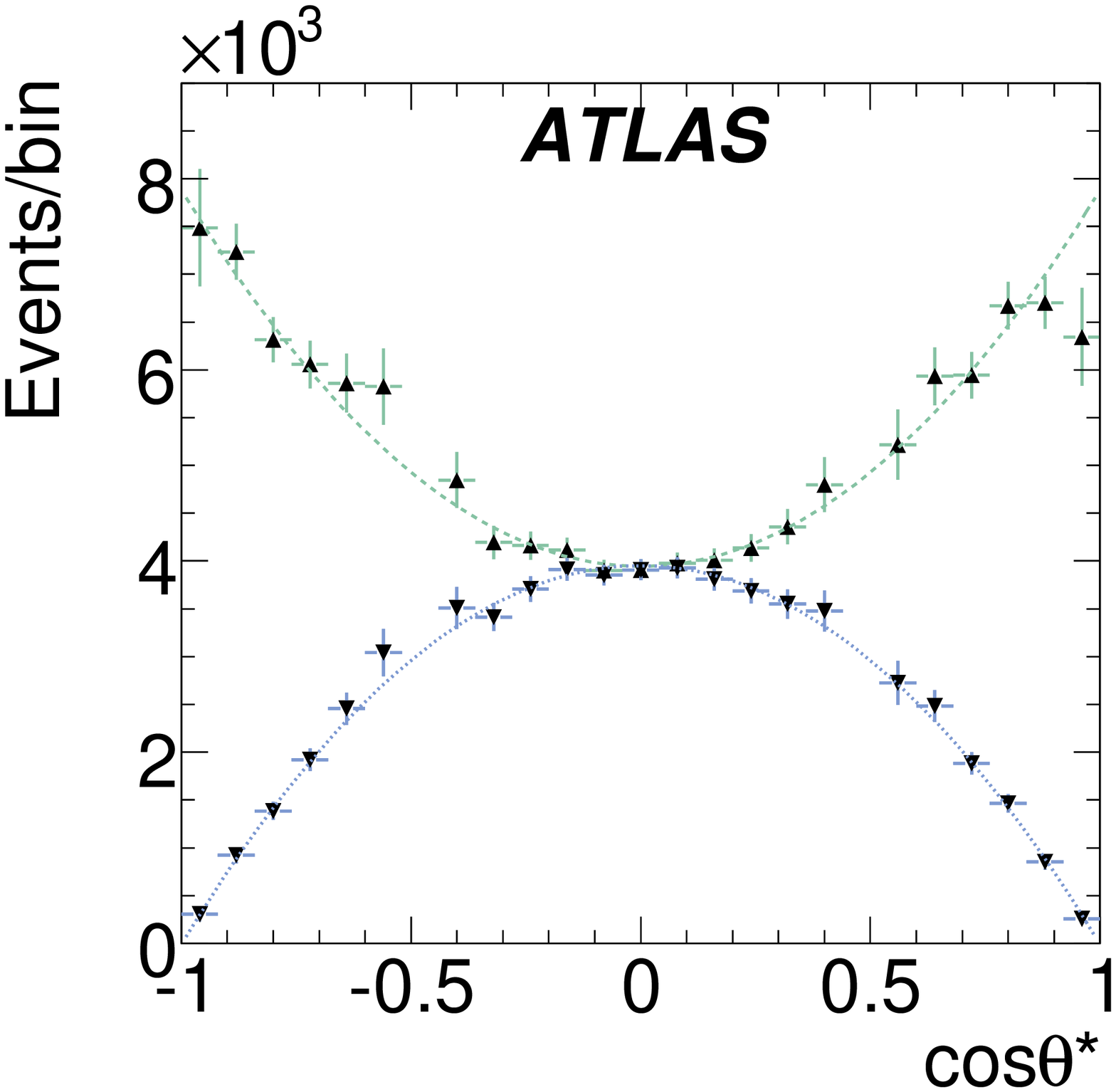}
      \vspace*{-0.3cm}%
      \caption{\label{fig:quarkoniapolarizationposneg}%
	Combined and corrected distributions in the polarization angle
	$\cos\theta^*$ for longitudinally (dotted lines) and
	transversely (dashed lines) polarized \jpsi\ mesons for the 
	13~GeV~$< p_{T,\jpsi} <$~15~GeV slice for 10~\ipb\ and
	the ATLAS study.
      }
    }
  \end{center}
\end{figure}


For the measurement of the prompt \jpsi\ and $\Upsilon$ cross sections,
the ATLAS  collaboration plans to use two methods. 
The first method relies on a di-muon
trigger ($p_{T,\mu} > 4$~GeV), 
applies offline cuts of 6 and 4~GeV $p_T$ to the two muon tracks
and requires them to orignate from the
primary vertex (pseudo-proper time $< 0.2$~ps).  
Within a mass window of $m_\jpsi^{PDG} \pm 300$~MeV 
($m_\Upsilon^{PDG} \pm 1$~GeV) about 150\,000 \jpsi\ (25\,000 $\Upsilon$)
are reconstructed with a signal over background ratio of $S/B = 60$ (10)
per 10~\ipb, 
see Fig.~\ref{fig:quarkoniamasses}.
The second method 
employs a single muon trigger ($p_{T,\mu} > 10$~GeV),
requires a track within a cone of $\Delta R < 3$ around the muon
direction and forces both to originate from the primary vertex.  About 
160\,000 \jpsi\ (20\,000 $\Upsilon$) with a lower 
$S/B = 1.2$ (0.05)
are selected per 10~\ipb.  
A combined measurement precision on the 1\% (5\%) level is estimated for
\dsd{}{\pt}\ for 10~\ipb\ of \jpsi\ ($\Upsilon$) events. 

The CMS collaboration plans similar measurements, relying on the
reconstruction of \jpsi\ mesons.  As an example for CMS's
capabilities\,\cite{CMSjpsi}, 
the extraction of $\jpsi$ mesons from 200\,000 Monte
Carlo $B_s\ra \jpsi(\mu^+\mu-)\,\phi$ events 
results in an overall Level~1 trigger plus
reconstruction efficiency of 10.1\%.
A \jpsi\ mass resolution of 34~MeV is achieved.



In order to extract the \jpsi\ and $\Upsilon$ polarizations 
$\alpha = (\sigma_T-2\sigma_L)/(\sigma_T+2\sigma_L)$ 
the simulated data are fitted to the acceptance-corrected $\cos\theta^*$
distributions, where $\theta^*$ specifies the polarization
angle of the $\mu^+$ w.r.t.\ the \jpsi\ momentum in the rest frame of the
\jpsi.  
Because of the different kinematic acceptance regions in $\cos\theta^*$ 
of the di-muon and single-muon trigger cuts, 
the sensitivity may improve when both samples are combined.
As shown in Fig.~\ref{fig:quarkoniapolarizationposneg}, the different
polarization states of the \jpsi\ can be distinguished well.  
With 10~\ipb\ of data, ATLAS estimates the statistical precision for the
measurement of $\alpha$ in \pt\ bins up to 20~GeV 
to be of order 0.02 -- 0.06 for \jpsi, depending
on the level of polarization itself, and of order 0.20 in the case of
the $\Upsilon$.


\section{Conclusions}

The ATLAS and CMS collaborations are preparing to measure the cross
sections for beauty and heavy quarkionia production at low \pt\ via
methods involving muonic decays or 
the $\ptrel$ method, and for higher 
\pt\ using $b$-tagging methods.  These and  
measurements of the polarization of \jpsi\ and $\Upsilon$ mesons 
will be performed 
with early data (10 to 100~\ipb).

\section{Acknowledgements}


We are indebted to our colleagues in ATLAS and CMS,
and the funding agencies supporting the experiments.  
We thank the Moriond~QCD team for organizing this excellent
conference, and are grateful to the German Ministry for Education and
Research (BMBF) for financial support.


\section{References}

\begin{thebibliography}{99}

\bibitem{kartvelishvili}
  {\it e.g.:}
  V. Kartvelishvili [ATLAS Collaboration],
  Nucl.\ Phys.\ Proc.\ Suppl.\ 164:161-168, 2007.

\bibitem{kramer-cdf}
  M.~Kr\"amer,
  Prog.\ Part.\ Nucl.\ Phys.\  {\bf 47} (2001) 141, 
  and references therein.

\bibitem{Abe:CDF}
  F.~Abe {\it et al.}  [CDF Collaboration],
  Phys.\ Rev.\ Lett.\  {\bf 69} (1992) 3704.



\bibitem{D0polar}
  V.~M.~Abazov {\it et al.}  [D0 Collaboration],
  D\O\, Note 5089-conf.

\bibitem{CMSbjet}
  V.P. Andreev {\it et al.} [CMS collaboration],
  CMS Note 2006/120.

\bibitem{CMSjpsi}
  Z. Yang and S. Qian [CMS collaboration],
  CMS Note 2007/017.

\end{thebibliography}
\end{document}